\documentstyle[amstex,12pt]{article}

\begin{document}

\setcounter{page}{0}
\thispagestyle{empty}

\noindent
\vspace{1.0cm}
\begin{center}
{\Large\bf A $q$-deformed Quantum Mechanics}
\end{center}
\vspace{1cm}
\begin{center}
{\bf Jian-zu Zhang$^{a,b,c,}$}
\end{center}
\vspace{1cm}
\begin{center}
\begin{description}
\item{a)} {Max-Planck-Institut f\"ur Physik (Werner-Heisenberg Institut),\\
F\"ohringer Ring 6, D-80805 M\"unchen, Germany}\\
\item{b)} {Theoretische Physik, Universit\"at Kaiserslautern,
PO Box 3049,\\
D-67653 Kaiserslautern, Germany}\\
\item{c)} {Institute for Theoretical Physics, Box 316, East China
University of \\
Science and Technology,
130 Mei Long Road, Shanghai 200237, P.R.China}\\
\end{description}
\end{center}
\vspace{2cm}

With a $q$-deformed quantum mechanical framework, features of the
uncertainty relation and a novel formulation of the
Schr\"odinger equation are considered.\\
\vspace{0.8cm}
\begin{description}
\item{\phantom{e)}}{PACS: 03.65.Bz}\\
\end{description}
\newpage

In searching for possible new physics at short distances (or high
energy scale) consideration of the space structure is a useful
guide. Quantum groups are a generalization of symmetry groups
which have been used successfully in physics. A general feature of
spaces carrying a quantum group structure is that they are
noncommutative and inherit a well-defined mathematical structure
from quantum group symmetries. In applications in physics,
questions arise whether the structure can be used for physics at
short distances and what phenomena could be linked to it.
Recently, starting from such a noncommutative space as
configuration space a generalization to a phase space is obtained
\cite{eins}. This noncommutative phase space is derived from the
noncommutative differential structure on configuration space
\cite{zwei}. Such noncommutative phase space is a $q$-deformation
of the quantum mechanical phase space and thus all the machinery
used in quantum mechanics can be applied in $q$-deformed quantum
mechanics \cite{eins,drei,vier,fuenf}.

In this letter we discuss the essential new features of a
$q$-deformed quantum mechanics: (i) A $q$-deformed uncertainty
relation: Here We find that the lowest limit of the Heisenberg
uncertainty is undercut. (ii) A $q$-deformed dynamical equation
which is found to be non-linear. The perturbative expansion of the
later shows complex structure. In the lowest order approximation
this equation is just the Schr\"odinger equation. The
characteristics of the new equation are essentially
non-perturbative. The qualitative behavior of its non-perturbative
solutions is different from that of the Schr\"odinger equation.
For example, the spectrum of the $q$-deformed harmonic
oscillator is exponentially spaced \cite{fuenf}.\\

{\bf 1. A $q$-deformed uncertainty relation}

\vspace{0.4cm}

The starting point of our investigation is the following
$q$-deformed Heisenberg algebra [1,6]:
\begin{equation}
q^{1/2}XP - q^{-1/2} PX = iU
\end{equation}
\begin{equation}
UX = q^{-1} XU, \quad UP = q PU
\end{equation}
with the conjugation properties
\begin{equation}
P^{\dagger} = P,\quad X^{\dagger} = X,\quad U^{\dagger} = U^{-1}.
\end{equation}

This $q$-deformed algebra is derived from the noncommutative
differential structure on configuration space \cite{zwei}.
However, if $X$ is assumed to be a hermitean operator in a Hilbert
space, the usual quantization rule $p\to -i\partial_X$ does not
yield a hermitean momentum operator \cite{eins}. In order to
define conjugation of $\partial_X$, and then a hermitean momentum
operator $P$, it is necessary to introduce a unitary scaling
operator $U$ satisfying (2)
\footnote{The definitions of the scaling operator $U$, the
conjugate $\bar \partial_X$ and $\partial_X$ and the hermitean
momentum operator $P$ are \cite{eins}:
$U^{-1}=q^{1/2}\left[1+(q-1)X\partial_X\right], \quad \bar
\partial_X=-q^{-1/2}U\partial_X, \quad
P=-\frac{i}{2}\left(\partial_X -\bar \partial_X\right).$}.
In Eqs. (1) and (2) the parameter $q$ is real and $q>1$. For the
case of $q=1$, the scaling operator $U$ is reduced to a unit
operator, and Eq.(1) reduces to the Heisenberg algebra. The
non-trivial properties of the operator $U$ lead to a richer
structure of algebra (1) and (2) than the Heisenberg algebra. From
(1) and (3) we obtain
\begin{equation}
XP - PX = iC
\end{equation}
where
\begin{equation}
C= (U + U^{-1})/(q^{1/2} + q^{-1/2}).
\end{equation}

In order to show the special characteristics of the $q$-deformed
uncertainty relation, we prove the following lemma.\\

Lemma. In any state the expectation value of the operator $C$ in
(5) satisfies
\begin{equation}
| < C > | \le 1.
\end{equation}

We note that $|<U+U^{-1}>|\le 2$. Because $q^{1/2}+q^{-1/2}\ge 2$
for any $q>0$, we obtain lemma (6).

Equation (4) gives
\begin{equation}
\Delta X\cdot\Delta P \ge {1\over 2} | <C>|.
\end{equation}
where $\Delta A=\sqrt{<(A-<A>)^2>}$. Lema (6) shows that the
lowest limit of the Heisenberg uncertainty relation $\Delta
X\cdot\Delta P=\frac{1}{2}$ is undercut. It is interesting to show
[7] that for the irreducible eigenstate $|n, \sigma>|s>$ of $P$
[1], $P|n, \sigma>|s>=P_0 |n, \sigma >|s>$ with $P_0=\sigma s q^n,
(\sigma = \pm 1; 0<s<1; n=0,1,2,...)$, we have $\Delta P=0$, but
$\Delta X$ is still finite. In fact, using (1)-(3), we obtain
$\Delta X=(q-q^{-1})^{-1}(q+q^{-1})^{1/2}P_0^{-1}$. Thus we
conclude that in the state $|n, \sigma >|s>, \Delta X\Delta P=0$.
This is a surprising qualitative deviation from Heisenberg's
uncertainty relation. It therefore raises the question where or
not the conventional uncertainty relation is recovered at large
scales. Unfortunately because of the complicated relations among
$X,$ $P$ and $U,$ an explicit form of the right-hand side of the
uncertainty relation as a function of $\Delta X$ and $\Delta P$ is
not obtained at this stage. Thus when $q$ is some fixed value not
equal to one it is still an important open question whether this
$q$-deformed quantum mechanics does at all reproduce ordinary
quantum mechanics at large scales. (7) may prove to be an
important result within this formulation of a $q$-deformed quantum
mechanics. Perhaps insights of a possible new physics just come
from here.

The issue of uncertainty relations in the context of quantum group
symmetric Heisenberg algebras was first considered by Kempf
\cite{Kempf}.

\vspace{0.4cm}

{\bf 2. The $q$-deformed dynamical equation}

\vspace{0.4cm}

The variables of the $q$-deformed algebra (1) and (2) can also be
expressed in terms of the variables of an undeformed algebra.
There are three pairs of canonical conjugate variables [1]: (i)
The variables $\hat x,\hat p$ of the undeformed quantum mechanics;
they satisfy $[\hat x, \hat p] = i$. (ii) The variables $\tilde x,
\tilde p;$ which are obtained by canonical transformation of $\hat
x$ and $\hat p$: ${\tilde p} = f({\hat z}){\hat p},\quad {\tilde
x} = {\hat x} f^{-1}({\hat z})$ where
\begin{equation}
f^{-1} ({\hat z}) =\frac{[{\hat z}-1/2]}{\hat z - 1/2}, \quad
{\hat z} = -\frac{i}{2}(\hat x\hat p + \hat p\hat x),
\end{equation}
and $[A] = (q^A-q^{-A})/(q-q^{-1})$ for any $A.$ The variables
$\tilde x, \tilde p$ also satisfy $[\tilde x, \hat p] = i$. (iii)
The $q$-deformed variables $X$ and $P$ where $X,$ $P$ and the
scaling operator $U$ are related to $\tilde x$ and $\tilde p$ in
the following way:
\begin{equation}
P = f^{-1}({\tilde z}) {\tilde p}, \quad X = {\tilde x}, \quad U =
q^{\tilde z}.
\end{equation}
in (9) $\tilde z$ and $f^{-1}(\tilde z)$ are defined by the same
equations (8) for $\hat z$ and $f^{-1}(\hat z)$. It is
easy to check that $X, P$ and $U$ in (9) satisfy (1)-(3).\\

our starting point is to use the $q$-deformed variables $X$ and
$P$ to write down the Hamiltonian, then using (9) to represent $X$
and $P$ by $\tilde x$ and $\tilde p$. Because of $[\tilde x,
\tilde p] = i$, (thus in the $\tilde x$ representation $\tilde p =
-i\tilde \partial,$ where $\tilde \partial=\partial/\partial
{\tilde x}$) all the machinery of quantum mechanics can be used
for the ($\tilde x, \tilde p$) system. The $q$-deformed
Hamiltonian of the system with potential $V(X)$ is $H(X,P) =
P^2/2m+ V(X).$ Using (1)-(3) and (9) the stationary dynamical
equation of $q$-deformed quantum mechanics reads as
\begin{equation}
\begin{split}
\{-{1\over2m}(q-q^{-1})^{-2}{\tilde x}^{-2}
[&q(q^{-2{\tilde x}
{\tilde\partial}}-1)+q^{-1}(q^{2{\tilde x}{\tilde\partial}}-1)] \\
&+V({\tilde x})\}\psi ({\tilde x}) = E_q\psi ({\tilde x}).
\end{split}
\end{equation}
Eq.~(10) is a non-linear equation which is a $q$-generalisation of
the Schr\"odinger equation.

For the case $q$ is close to 1, we let $q=e^f, 0<f\ll 1$. The
perturbative expansion of the Hamiltonian is then
\begin{eqnarray}
H=&-&\frac{1}{2m}(2f+\frac{1}{3}f^3+\dots)^{-2}{\tilde x}^{-2}
[4f^2{\tilde x}^2{\tilde\partial}^2+\frac{1}{3}f^4(4{\tilde x}^4
{\tilde\partial}^4\nonumber\\
&+&16{\tilde x}^3{\tilde\partial}^3+10{\tilde
x}^2{\tilde\partial}^2)+ \dots ]+V({\tilde x}).
\end{eqnarray}

Thus to the lowest order in $f$, Eq.~(10) reduces

\begin{equation}
[-\frac{1}{2m}{\tilde\partial}^2+V({\tilde x})]\psi ({\tilde x}) =
E\psi ({\tilde x}).
\end{equation}

This is just the Schr\"odinger equation for the ($\tilde x, \tilde
p$) system. In (11) the next order correction of $H$ shows a
complex structure which amounts to some additional momentum
dependent interaction.

In the above we constructed the $q$-deformed Hamiltonian of the
variables $X$ and $P$ in analogy with the undeformed system.
Another possible way to construct a $q$-deformed Hamiltonian is
that the system should act in accordance with a special algebra.
An example
is a $q$-deformed harmonic oscillator \cite{fuenf}.\\

If $q$-deformed quantum mechanics is a correct theory, its
corrections to the undeformed theories must be very small at the
energy range which can be reached by present-day experiments. In
view of the present accuracy of tests of quantum electrodynamics
at least down to $10^{-17}$ cm, the effects of $q$-deformed
quantum mechanics would show up at distances much smaller than
$10^{-17}$ cm. We hope that the $q$-deformed uncertainty relation
might show some evidence in present-day
experiments.\\

{\bf Acknowledgements}

\vspace{0.4cm}

The author would like to thank Prof. J. Wess very much for giving
the author the opportunity to join his project on quantum groups
and for many stimulating helpful discussions and Prof. H. J. W.
M\"uller-Kirsten for many helpful comments. He would also like to
thank the Max-Planck-Institut f\"ur Physik
(Werner-Heisenberg-Institut) for financial support and the Sektion
Physik, Universit\"at M\"unchen, for warm hospitality. His work
has also been supported by the Deutsche Forschungsgemeinschaft
(Germany), the National Natural Science Foundation of China under
Grant No. 19674014, and the Shanghai Education Development
Foundation.


\end{document}